# Understanding Computational Science and Engineering (CSE) and Domain Science Skills Development in National Laboratory Postgraduate Internships

Morgan M. Fong,[1, 2] Hilary Egan,[1] Marc Day,[1] Kristin Potter,[1] Michael J. Martin[1]

[1]National Renewable Energy Laboratory, Golden, CO

[2]University of Illinois, Urbana-Champaign, IL. Current Address: University of Texas, Austin, TX

**Abstract**

*Background:* Harnessing advanced computing for scientific discovery and technological innovation demands scientists and engineers well-versed in both domain science and computational science and engineering (CSE). However, few universities provide access to both integrated domain science/CSE cross-training and Top-500 High-Performance Computing (HPC) facilities. National laboratories offer internship opportunities capable of developing these skills.

*Purpose:* This study presents an evaluation of federally-funded postgraduate internship outcomes in computational science at a national laboratory. This study seeks to answer three questions: 1) What computational skills, research skills, and professional skills do students improve through internships at the selected national laboratory? 2) Do students gain knowledge in domain science topics through their internships? 3) Do students' career interests change after these internships?

*Design/Method:* We developed a survey and collected responses from past participants of five federally-funded internship programs and compare participant ratings of their prior experience to their internship experience.

*Findings:* Our results indicate that participants improve CSE skills and domain science knowledge, and are more interested in working at national labs. Participants go on to degree programs and positions in relevant domain science topics after their internships.

*Conclusions:* We show that national laboratory internships are an opportunity for students to build CSE skills that may not be available at all institutions. We also show a growth in domain science skills during their internships through direct exposure to research topics. The survey instrument and approach used may be adapted to other studies to measure the impact of postgraduate internships in multiple disciplines and internship settings.

## I. INTRODUCTION & LITERATURE REVIEW

Scientific discovery and technological advancement increasingly depend upon the use of advanced computing, including modeling and simulation (mod/sim), data science, and machine learning and artificial intelligence (ML/AI). [Rude et al., 2018], [Peters et al., 2024], [Shiflet, 2002] The development and application of these tools has emerged as its own scientific discipline, *computational science and engineering, or CSE* [Rude et al., 2018]. In many cases, the most effective use of advanced computing for scientific discovery and technology development occurs when students are cross-trained in both CSE and *domain science* areas such as engineering, physics, chemistry, or biology [McInnes et al., 2023]. The value of these skills increases when scientists and engineers are able to make effective use of high performance computing (HPC) systems which are characterized by increasing complexity, including massive parallelization, heterogeneous architectures, and the use of complex programming models [Evans et al., 2022].

The need for CSE and HPC expertise is not confined to academic and research institutions. Industry's growing reliance on CSE and HPC is evident in the major investments in world-class supercomputers by [TOP500.org, 2024] and accelerating adoption of cloud-based HPC [Netto et al., 2018]. The ability to effectively use commercial computational software, and scientific programming skills in languages such as C++, Python, and MATLAB have been identified as highly-sought after skills for engineers at the post-bachelor's, -master's, and -doctoral levels through analysis of job advertisements [Fleming et al., 2024] and surveys of PhD earners employed in industry [Watson and Lyons, 2011]. Furthermore, both programming [Fleming et al., 2024] and HPC [Snapp-Childs et al., 2025] skills are associated with significant salary premium for engineers and scientists pursuing careers in industry.

In spite of increasing demand for domain scientists trained in CSE skills [McInnes et al., 2023], [Tillotson, 2017], there are limits to the ability of academic institutions to provide both the expertise and hardware needed for effective training [Kejriwal and Petryshyn, 2024]. Training may be limited by lack of access to hardware, with many universities only able to support small-scale systems (e.g., [Verbic et al., 2017], [Ochs and Miller, 2015]). As a result, relatively few university students have access to modern supercomputers, which often have vastly different capabilities and architectures than those of smaller, more affordable clusters. Based on the November 2024 ranking of the top 500 most powerful supercomputers in the world [TOP500.org, 2024], 173 supercomputers are located in the United States (US), the context in which this study is situated. Of those in the US, 67 are hosted in government research facilities, including 38 at Department of Energy (DOE) National Laboratories, 3 are hosted in research foundations, and 80 are hosted in industry and vendors. However, only 23 top-500 supercomputers are hosted by US universities. These computers are concentrated in 17 very high research activity universities [Indiana University Center for Postsecondary Research, 2021] located in 9 states, further limiting access to hands-on training.

Even in universities with access to larger-scale compute facilities, integrating domain sciences with CSE training faces institutional challenges. These include the constraints of existing curriculum requirements (e.g., [Kejriwal and Petryshyn, 2024]), lack of evaluation of learning outcomes (e.g., [Wilson and Dey, 2016]), or inadvertently de-emphasizing computational skills when integrating domain science topics (e.g., [Duarte et al., 2020], [Angelaki et al., 2024]). These challenges are often exacerbated by a shortage of faculty in key areas of CSE. [Zwetsloot and Corrigan, 2022]

Making national laboratory computing resources and staff expertise available to university students is a possible solution to these challenges. Multiple strategies exist. Because DOE computing resources are used by academic research groups across the United States, many research groups have developed their own internal documentation and training for the systems that they have access to, and the tools that are necessary of support their work. A second related approach is to provide national laboratory computing time and staff support directly to university partnerships seeking to provide training at the institutional level. [ALCF, 2024], [ORNL, 2024], [Sanchez, 2024].

A third approach is to integrate students into national laboratory research that uses HPC through internships. Because national laboratories that host HPC facilities also host communities of researchers using HPC, students are likely to have access to a broad range of expertise in both CSE and domain science.

Previous assessments of national laboratory internships not specifically focused on CSE shows multiple impacts, including:

- Increased retention in science, technology, engineering, and mathematics (STEM) careers. [Cote et al., 2025], [Foltz et al., 2011]

- Increased research skills. [Foltz et al., 2011], [Cote et al., 2025]
- Increased professional networks. [Cote et al., 2025].

Internships in the broader field of computer science are linked multiple benefits, including:

- Development of technical skills. [Kang and Girouard, 2022], [Minnes et al., 2021], [Kapoor and Gardner-McCune, 2019], [Shiflet, 2002]
- Opportunities for career exploration. [Kang and Girouard, 2022], [Kramarczuk et al., 2024], [Kapoor and Gardner-McCune, 2019]
- Expanded professional networks that contribute to future professional success. [Kramarczuk et al., 2024], [Minnes et al., 2021], [Ferguson et al., 2024]

However, the question of the development of specific technical skills in national laboratory internships, specifically in CSE and HPC, remains open. While there is substantial literature on training in HPC, these papers tend to focus on descriptions of internships at academic computing centers [Jacobs et al., 2014] or workshops designed for specific supercomputer sites (e.g., [Connor et al., 2016], [Dubey et al., 2020], [Bautista and Sukhija, 2024], [Wilson and Dey, 2016]). The potential benefits of internships in the national laboratory environments, which combine expertise in CSE, domain science, and HPC operations has not been examined.

Our study aims to address this gap through an evaluation of federally funded, postgraduate (including postbaccalaureate-, master's-, and doctoral-level) internships hosted at the National Renewable Energy Laboratory (NREL). To this end, we ask the following research questions:

*Research question 1:* What computational skills, research skills, and professional skills do postgraduate students improve through their internships at NREL?

*Research question 2:* What domain-science topics do postgraduate students gain more familiarity with through their internships at NREL?

*Research question 3:* Do postgraduate students' career interests change after their internships at NREL?

To answer these questions, we conducted a survey study with past participants of five federally funded internship programs hosted at NREL. These programs serve post-baccalaureate and graduate participants from a variety of disciplines and educational institutions. While this survey is specifically focused on the intersection of CSE and NREL's core research, results are likely to be broadly applicable to interdisciplinary training that combines CSE with other domain sciences.

We provide additional context of the internship programs and internship site in Section II. Section III-A details the methods used to design the survey as well as collect and analyze survey data. Section V includes discussion of our results, conclusions, and possibilities for future work.

## II. OVERVIEW OF HOST INSTITUTIONS AND INTERNSHIP PROGRAMS

This study examines the impact of *federally-funded internships* in CSE at the National Renewable Energy Laboratory (NREL), a Department of Energy (DOE) National Laboratory located in Golden, Colorado. The laboratory has additional campuses in Arvada, Colorado, Fairbanks, Alaska, and Washington, DC. As of 2024, the laboratory had 3,675 staff, including regular employees, postdoctoral researchers, visiting professionals, and subcontractors.

NREL's research focuses on energy systems and energy efficiency in support of the research objectives of multiple DOE offices, including the Office of Energy Efficiency and Renewable Energy (EERE), the Office of Electricity, the Office of Fossil Energy and Carbon Management, and the Office of Science. NREL increasingly uses computational science tools, including mod/sim, data science, and ML/AI to perform this research. NREL's efforts in these areas are led by the Computational Science Center (CSC). CSC hosts

Kestrel, NREL's third generation of High Performance Computing (HPC) that is available to all researchers funded by EERE. The center also engages in research in computational science spanning fundamental work on algorithms, software development, scientific visualization, and application to a range of topics in energy systems [Martin et al., 2025]. As of November 2024, CSC staff includes 7 managers, 11 administrative staff, 65 scientific staff, 44 technical staff (responsible for operation of the HPC facilities), and 14 postdocs, or a total of 141 staff. Staff expertise covers a broad range of CSE topics, including modeling and simulation, data science and visualization, and machine learning and artificial intelligence. Staff expertise also includes software engineering, systems administration, and data center infrastructure management and facility integration.

NREL hosts interns in CSE through two mechanisms: federal educational fellowships, with financial support from the fellowship sponsor, and as temporary staff working on specific projects, with financial support from project funds. Federal educational fellowships specify that the student be achieving specific educational objectives while contributing to active research projects. Both types of internships are available to undergraduate, postbaccalaureate, and graduate students. Because CSE is often an advanced undergraduate or graduate-level topic, NREL's internships in this area are heavily weighted towards postbaccalaureate and graduate students. Based on this, the current study focuses on postgraduate students supported through federal educational fellowships.

NREL currently brings postbaccalaureate and graduate student interns in CSE to the laboratory through five federally funded educational internship programs. Four of these programs are funded by the DOE: the Computational Science Graduate Fellowship (CSGF), the Science Undergraduate Laboratory Internship (SULI) program, the Office of Energy Efficiency and Renewable Energy (EERE) High Performance Computing for Energy Innovation Internship program (HPC4EI), and the National Nuclear Security Administration's Minority Serving Institutions Internship (NNSA-MSIIP). A fifth program is funded by the National Science Foundation (NSF): the Mathematical Science Graduate Internship (NSF-MSGI).

For all of these programs, each intern is assigned a mentor (i.e., a staff member at the national lab), and each mentor is expected to submit a plan to the sponsor that identifies learning objectives as well as scientific objectives for the internship. Once the internship begins, interns have the opportunity to learn from direct interaction with mentors and other members of the technical staff. In addition to this mentorship, NREL's HPC operations staff offers a broad range of structured trainings and tutorials on topics in HPC, as well as providing one-on-one advising when needed. Each of these programs has unique goals, eligibility criteria, and post-internship report requirements, which we describe next.

DOE CSGF is a highly competitive four-year fellowship for doctoral students in science and engineering who incorporate computational science into their work [Krell Institute, 2024]. Applicants must be US citizens or permanent residents. As part of their fellowship, students must spend at least one semester at one of 21 DOE facilities pursuing a project outside of their dissertation work. The 20-30 students seeking placement each year work directly with practicum coordinators at different facilities to identify a facility, mentor, and project that meet their interests. Students must complete an initial practicum within the first two years of their fellowship and are free to participate in second or third practicums, either with the same DOE facility and mentor, or a different DOE facility and/or mentor. CSC manages placement within NREL, though assigned mentors and projects can within CSC or in another center with computationally intensive projects. While placements are generally made during the summer, they are also occasionally made in the fall or spring semesters.

DOE SULI is a one-semester internship program open to US citizens or permanent residents who are undergraduates, postbaccalaureate, or graduate students in their first two years since completing their bachelor's degree [DOE Office of Science, 2024]. Students are not required to be majoring in a STEM field but must have at least 6 credit hours in STEM and a minimum 3.0 out of 4.0 grade point average. SULI students are eligible for placement at 17 different DOE facilities. Students may complete up to two placements, either with the same DOE facility and mentor, or a different DOE facility and/or mentor. Students may be placed for the summer, fall, or spring semesters.

The EERE HPC4EI internship program is linked to the DOE's HPC4EI program [Department of Energy, 2024]. The HPC4EI program matches DOE national laboratory computing capabilities with industrial partners to solve key challenges in industrial energy use and the manufacture of energy materials. As part of its industrial competitiveness mission, HPC4EI offers internships to US citizens in HPC-relevant fields as undergraduate, postbaccalaureate, and graduate students [ORAU, 2023]. HPC4EI interns work directly on projects sponsored by the HPC4EI program.

NNSA-MSIIP targets building research capacity at Minority Serving Intuitions (MSIs) as defined by the US Department of Education [U.S. Department of Education, 2024]. NNSA-MSIIP is open to US citizens of all backgrounds, academic levels, and majors at these institutions [NNSA and ORAU, 2024]. Students can complete an in-person or virtual internship and have the option of extending into the academic year.

NSF-MSGI is open to doctoral students in mathematics, statistics, and applied mathematics. The goal of NSF-MSGI is to expose these students to non-academic careers [NSF, 2024]. Unlike the other programs described, NSF-MSGI is open to international students. Students may complete multiple internships, but each internship must be with a different mentor.

All of these programs operated as fully virtual programs in 2020 and 2021 due to the COVID pandemic. DOE CSGF deferred all summer 2020 placements to summer 2021 placements, but did allow virtual placements in fall of 2020 and spring of 2021. Neither EERE HPC4EI nor NSF-MSGI operated in summer of 2024 due to budget uncertainties.

## III. METHODS

### *A. Survey Design*

The research team collaboratively designed the survey. Given the purpose of the survey to understand interns' 1) technical and professional skill development and 2) pathways to future careers, survey questions were organized and designed around the following themes: Computational Skills, Familiarity with NREL-relevant domain science topics such as energy generation and efficiency, Research Skills, Professional Skills, and Career Interests. Full questions for each theme are provided in Table I.

For each theme, the research team identified relevant skills and topics based on their extensive experience with the internship programs described in Section II. More specifically, four of the co-authors (Martin, Egan, Day, and Potter) have extensive involvement in NREL's internship programs. Michael Martin has served as NREL's coordinator for the DOE CSGF, EERE HPC4EI, and NNSA-MSIIP programs as well as mentored students through the NSF-MSGI program. Hilary Egan has served as NREL's co-coordinator for the DOE CSGF program and has mentored students through the DOE SULI program. Marc Day has mentored students through the DOE CSGF and NSF-MSGI programs. Kristi Potter has mentored students through the DOE CSGF and SULI programs.

For the questions related to computational skills, inspiration was taken from a previous report on CSE education [Rude et al., 2018]. The research team iteratively adjusted the questions for clarity and understanding across each internship program's context.

Additionally, optional open-ended response options were provided for participants to describe other computational skills, research skills, professional skills, and topics that they improved or gained familiarity with during their internship to capture areas we did not explicitly ask about.

# Table I. Survey Questions Organized by Theme

*All questions were asked with respect to before the internship and during the internship. Questions in career interest were split by domain and sector.*

<table>
<tr><th>Theme</th><th>Question Context, Response Options</th><th>Skills and Topics</th></tr>
<tr><td rowspan="2">Computational Skills</td><td>Prior to participating in [internship program], how much experience did you have with the following? <i>No experience, Some experience (e.g., 1-2 projects), Extensive experience (e.g., took advanced coursework), Expert (e.g., focus of degree, years of experience)</i></td><td rowspan="2">1. Writing general-purpose programs<br>2. Writing programs for scientific problems<br>3. Writing distributed or parallel programs<br>4. Writing software collaboratively<br>5. Performing data visualization and analysis<br>6. Using AI/ML to develop models<br>7. Developing simulations of systems or phenomena (e.g., computational fluid dynamics, computational chemistry)<br>8. Using high performance computers or supercomputers<br>9. Using software libraries or open-source code</td></tr>
<tr><td>During your internship at NREL, to what extent did you improve your skills in the following? <i>No improvement, Some improvement, Moderate improvement, Substantial improvement</i></td></tr>
<tr><td rowspan="2">Familiarity with Sustainability and Renewable Energy Topics</td><td>Prior to participating in [internship program], how familiar were you with the following? <i>No familiarity, Somewhat familiar (e.g., news, social media), Very familiar (e.g., took advanced coursework) Extremely familiar (e.g., focus of degree)</i></td><td rowspan="2">1. Renewable energy<br>2. Energy efficiency<br>3. Materials usage<br>4. Circular economies<br>5. Environmental justice<br>6. Energy equity<br>7. Sustainable Computing</td></tr>
<tr><td>To what extent did your internship at NREL increase your familiarity with the following topics? <i>No increase in familiarity, Some increase in familiarity, Moderate increase in familiarity, Substantial increase in familiarity</i></td></tr>
<tr><td rowspan="2">Research Skills</td><td>Prior to participating in [internship program] how much experience did you have with the following? <i>No experience, Once, 2-3 times, 4 or more times</i></td><td rowspan="2">1. Interview for a research position<br>2. Contribute to a research project<br>3. Work in a team with varying types of expertise<br>4. Author or co-author a research publication (e.g., journal article, conference paper, book chapter)</td></tr>
<tr><td>During your internship at NREL, how much experience did you have with the following? <i>No experience, Once, 2-3 times, 4 or more times</i></td></tr>
<tr><td rowspan="2">Professional Skills</td><td>Prior to participating in [internship program], to what extent would you agree or disagree with the following statements? <i>Disagree, Somewhat disagree, Somewhat agree, 4: Agree</i></td><td rowspan="2">1. I'm part of a research community where I can get feedback on my work<br>2. I can create an effective oral research presentation<br>3. I can design an informative research poster<br>4. I can discuss research findings with a technical audience<br>5. I can discuss research findings with a non-technical audience</td></tr>
<tr><td>During your internship at NREL, to what extent do you agree or disagree with the following statements? <i>Disagree, Somewhat disagree, Somewhat agree, 4: Agree</i></td></tr>
<tr><td rowspan="2">Career Interests</td><td>Prior to participating in [internship program], how interested were you in a career in the following? <i>Not interested, Somewhat interested Very interested, Extremely interested</i></td><td>1. Sustainability<br>2. Renewable energy<br>3. Another domain other than sustainability or renewable energy</td></tr>
<tr><td>As a result of your internship at NREL, how interested were you in a career in the following? <i>Not interested, Somewhat interested, Very interested, Extremely interested</i></td><td>1. Industry<br>2. Academia<br>3. National Renewable Energy<br>4. Other Department of Energy National Laboratory<br>5. Other federal research facility<br>6. Non-Profit<br>7. Another sector other than the above</td></tr>
<tr><td>If currently in Different Degree Program, Looking for Employment, or Currently Employed</td><td>To what extent is your [field of study/positions applying for/ current position] related to sustainability or renewable energy? <i>Not related, Somewhat related, Very related, Extremely related</i><br>How useful are the skills you improved at NREL in applying for [degree programs/positions]? <i>Not useful, Somewhat useful, Very useful, Extremely useful</i></td><td>1. Computational Skills<br>2. Familiarity with Sustainability and Renewable Energy Topics<br>3. Research and Professional Skills</td></tr>
<tr><td>If currently in the Same Degree Program, Different Degree Program, or Currently Employed</td><td>How often do you use the skills improved during your internship at NREL in your [current position]? <i>Never, Some of the time, Most of the time, All of the time</i></td><td>1. Computational Skills<br>2. Familiarity with Sustainability and Renewable Energy Topics<br>3. Research and Professional Skills</td></tr>
</table>

*B. Participant Background Information*

Participants were also asked to self-report the following academic background information:

- Prior experience at other Department of Energy National Labs (i.e., no experience, once, 2-3 times, or 4 or more times)
- Prior experience at other federal laboratories (i.e., no experience, once, 2-3 times, or 4 or more times)
- Year of first internship at NREL, as some participants may have extended their internship or returned to NREL at a later time
- Academic stage and major / field of study during internship at NREL or, for participants who were not enrolled in a degree program during their internship, major / field of study during degree prior to internship
- Location of internship (i.e., in person, hybrid, remote)
- Focus of internship on domain science topics (i.e., not at all, somewhat focused, very focused, or extremely focused)
- Current position at time of completing survey (i.e., in the same or different degree program, looking for employment, or currently employed)

Additional questions were asked based on participants' current positions to understand how their internship experiences may or may not have carried beyond their time at NREL. These questions are also provided in Table I.

*C. Participants*

We recruited participants from five federally-funded internships at NREL as described in Section II. Because these internship programs are open to a broader population than our intended focus, we applied the following inclusion criteria:

- Participants must have completed a bachelor's degree or equivalent prior to starting their internship at NREL
- Participants must have started their internship at NREL in 2018 or later
- Participants incorporated CSE into their internships, as evidenced by (1) the nature of the internship program (CSGF, HPC4EI, and NSF-MSGI), or, (2) for SULI and NNSA-MSIIP students, either (a) a mentor in CSC, or (b) use of NREL's supercomputer during their internship in another NREL center.

Many potential participants were identified given the research team's extensive involvement with the internship programs of interest. Public records of some internship programs were cross-referenced with internal NREL records to identify additional participants. Additionally, we contacted NSF-MSGI administrators for the list of past NSF-MSGI participants hosted at NREL. All study procedures follow our IRB-approved protocol (#DOE001036).

*D. Data Collection*

Participants were sent either a recruitment email or LinkedIn message in September 2024. The survey accepted responses for three weeks.

Because the authors have extensive experience working with these internship programs (see Section III-A for more details), Morgan Fong, was recruited as a graduate student intern to the research team to manage distribution of the survey and removal of personally identifiable information from responses. Martin, Egan, Day, and Potter only had access to de-identified and aggregated data.

### *E. Data Analysis*

All Likert-scale questions, except for questions related to Professional Skills, are coded on a scale of zero (i.e., "No experience," "No improvement," "No familiarity," "No increase in familiarity," "Not interested," "Not at all," "Not useful," "Not related,") to three (i.e., "Expert," "Substantial Improvement," "4 or more times," "Extremely familiar," "Substantial increase in familiarity," "Extremely interested," "Extremely focused," "Extremely related," "Extremely useful"). Professional Skills were coded on a scale of 1 (i.e., "Disagree") to 4 (i.e., "Agree").

For professional skills and career interests, one-sided, paired samples t-tests were used to determine if participant ratings after internships are greater than their rating from before internships. For computational skills, research skills, familiarity with domain science topics, one-sided, one sample t-tests were used to determine if participants reported at least "Some improvement" for computational skills, one experience for research skills during the internship, "Some increase in familiarity" for familiarity with domain science topics. In other words, the t-tests for computational skills, research skills, and familiarity with domain science topics determined if participants' post-internship ratings were > 1.

Similarly, one-sided one sample t-tests were used to determine if participants current positions are at least "Somewhat related" to NREL domain science topics, if their skills are at least "Somewhat useful" in applications, and if their skills are used at least "Some of the time." $\alpha < 0.05$ served as the cutoff for statistical significance.

### *F. Limitations*

The inclusion of a variety of internship programs provided a broader participant pool, but the present study remains limited to a small sample size. Additionally, the sample is subject to self-selection bias based on the selectiveness of the internship programs and who chose to respond to the survey. However, we are encouraged to see relatively high response rates across programs that indicate we may indeed have a representative sample of the intended population.

For participants who completed their internships less recently, their responses may be less accurate. We attempted to combat this by ensuring our response options were specific as possible to help participants select the most accurate response.

## IV. RESULTS

The total number of interns by program as well as corresponding response rates are included in Table II. Participants' background data are included in Table III. For anonymity purposes, the following categories are grouped together: participants who reported being in the first or second year of their PhD during their internship, participants who reported being in the third year of their PhD or later during their internship. Additionally, participants' fields of study were grouped as follows: science (e.g., physical, life, and biological sciences), engineering (e.g., chemical and mechanical engineering), mathematics (e.g., math, applied math, statistics), computing (e.g., computer science, computer engineering), and interdisciplinary (e.g., fields of study that included more than one of the above).

Three participants reported at least one prior experience at a DOE National Laboratory, and no participants reported prior experience at non-DOE federal research facilities.

On average, participants reported their internships were focused on domain science topics ($\mu = 1.46, \sigma = 0.83$). At the time of their internships, participants were studying a variety of domain sciences and mathematics. Averages and standard deviations for the main survey questions are included in Table IV. Next, we describe results by theme.

Table II. Total number of interns meeting inclusion criteria by program, year of internship and response rates.

*NO indicates that the program was non-operational for that year. NP indicates the program was operational, but NREL had no participation for that year.*

| Year | DOE CSGF | DOE SULI | NSF MSGI | EERE HPC4EI | NNSA MSIIP | Total |
|---|---|---|---|---|---|---|
| 2018 | 2 | 3 | 1 | NO | NO | 6 |
| 2019 | 3 | 1 | 3 | NO | NO | 7 |
| 2020 | 1 | 5 | 0 | 0 | NP | 6 |
| 2021 | 3 | 4 | 2 | 1 | NP | 10 |
| 2022 | 7 | 2 | 1 | 2 | NP | 12 |
| 2023 | 2 | 2 | 6 | 0 | NP | 10 |
| 2024 | 5 | 0 | NO | NO | 2 | 7 |
| Total | 23 | 17 | 13 | 3 | 2 | 58 |
| Response Rate | 52% | 29% | 38% | 67% | 0% | 41% |

Table III. Participants' academic backgrounds and internship location.

| | | |
|---|---|---|
| Academic Stage During Internship | Post-Baccalaureate | 3 |
| | Master's | 3 |
| | Early Stage (1st or 2nd year) PhD | 13 |
| | Late Stage (3rd year or later) PhD | 5 |
| Major or Field of Study | Science | 4 |
| | Engineering | 5 |
| | Mathematics | 5 |
| | Computing | 3 |
| | Interdisciplinary | 7 |
| Internship Location | In-person | 15 |
| | Hybrid | 4 |
| | Remote/Online | 5 |

### *A. Computational Skills*

On average, participants enter their internships with at least some experience across most computational skills. Additionally, participants report at least "some improvement" in general-purpose programming ($t(23) = 4.65, p = .000$), scientific programming ($t(23) = 4.66, p < .000$), data visualization & analysis ($t(23) = 5.17, p = .000$), use of HPC and supercomputers ($t(23) = 2.53, p = .009$), and using software libraries or open-source code ($t(23) = 5.35, p = .000$). The distribution of responses is provided in Figure 1.

While the improvements were not statistically significant across all skills, we note that 21 participants reported "moderate improvement" or "substantial improvement" in multiple skills. The open-ended responses mentioned additional specific skills such as MPI (Message Passing Interface) and manifold optimizations.

### *B. Familiarity with Domain Science topics*

On average, participants enter their internships with some familiarity with the domain science topics of energy systems and energy efficiency, and little to no familiarity with other topics. Participants report at least "some increase" in familiarity in renewable energy ($t(23) = 5.36, p = .000$) and energy efficiency ($t(23) = 1.89, p = .035$). The distribution of responses is provided in Figure 2.

While the increases in familiarity were not statistically significant across all skills, we note that six participants reported "moderate increase" or "substantial increase" in one topic, and 13 participants reported "moderate increase" or "substantial increase" in multiple topics.

Table IV. Survey questions organized by theme. All questions were asked with respect to before the internship and during the internship

*Questions on career interests were split by domain and sector.*

| Question | Pre-Internship $\mu(\sigma)$ | Post-Internship $\mu(\sigma)$ | t-test |
|---|---|---|---|
| Computational 1 | 1.54(0.78) | 1.67(0.70) | 4.65, p = 0.000 |
| Computational 2 | 1.67(0.92) | 1.79(0.83) | 4.66, p = 0.000 |
| Computational 3 | 0.71(0.69) | 0.79(0.98) | -1.04, p = 0.846 |
| Computational 4 | 0.71(0.81) | 1.08(1.06) | 0.39, p = 0.352 |
| Computational 5 | 1.50(0.72) | 1.96(0.91) | 5.17, p = 0.000 |
| Computational 6 | 0.62(0.77) | 1.00(1.22) | 0.00, p = 0.500 |
| Computational 7 | 1.29(1.00) | 1.29(1.20) | 1.19, p = 0.122 |
| Computational 8 | 1.21(0.93) | 1.62(1.21) | 2.53, p = 0.009 |
| Computational 9 | 1.62(0.97) | 2.04(0.95) | 5.35, p = 0.000 |
| Familiarity 1 | 1.25(0.74) | 1.83(0.76) | 5.36, p = 0.000 |
| Familiarity 2 | 1.21(0.66) | 1.38(0.97) | 1.89, p = 0.035 |
| Familiarity 3 | 0.75(0.68) | 0.88(0.99) | -0.62, p = 0.728 |
| Familiarity 4 | 0.38(0.49) | 0.50(0.83) | -2.94, p = 0.996 |
| Familiarity 5 | 0.79(0.66) | 0.62(0.82) | -2.23, p = 0.982 |
| Familiarity 6 | 0.67(0.64) | 0.54(0.83) | -2.70, p = 0.994 |
| Familiarity 7 | 0.67(0.76) | 0.88(0.90) | -0.68, p = 0.749 |
| Research 1 | 0.79(0.78) | 0.38(0.65) | -4.73, p = 1.000 |
| Research 2 | 1.96(0.75) | 1.50(0.83) | 2.94, p = 0.004 |
| Research 3 | 1.75(1.11) | 1.46(0.93) | 2.41, p = 0.012 |
| Research 4 | 1.33(1.01) | 0.88(0.85) | -0.72, p = 0.761 |
| Professional 1 | 3.33(0.70) | 3.67(0.56) | -1.88, p = 0.036 |
| Professional 2 | 3.58(0.50) | 3.46(0.98) | 0.68, p = 0.749 |
| Professional 3 | 3.42(0.58) | 3.04(1.27) | 1.33, p = 0.902 |
| Professional 4 | 3.38(0.65) | 3.54(0.83) | -0.75, p = 0.231 |
| Professional 5 | 3.17(0.70) | 3.00(0.98) | 0.64, p = 0.736 |
| Domain 1 | 1.54(1.02) | 1.92(0.88) | -2.39, p = 0.013 |
| Domain 2 | 1.62(1.01) | 2.25(0.79) | -3.98, p = 0.000 |
| Domain 3 | 1.38(0.77) | 1.71(0.62) | -3.39, p = 0.001 |
| Sector 1 | 1.29(0.55) | 1.17(0.70) | 0.83, p = 0.792 |
| Sector 2 | 1.12(1.03) | 1.08(0.97) | 0.24, p = 0.593 |
| Sector 3 | 1.54(0.78) | 2.42(0.78) | -4.14, p = 0.000 |
| Sector 4 | 1.54(0.83) | 1.83(0.87) | -2.60, p = 0.008 |
| Sector 5 | 1.25(0.79) | 1.29(0.75) | -0.27, p = 0.394 |
| Sector 6 | 0.67(0.70) | 0.67(0.87) | 0.00, p = 0.500 |
| Sector 7 | 0.38(0.58) | 0.46(0.59) | -0.81, p = 0.213 |
| Computational* | 2.40(0.74) | | 7.36, p = 0.000 |
| Familiarity* | 1.93(1.03 | | 3.50, p = 0.002 |
| Research & Professional* | 2.40(0.74) | | 7.36, p = 0.000 |
| Computational+ | 2.26(0.75) | | 8.04, p = 0.000 |
| Familiarity+ | 1.48(1.04) | | 2.21, p = 0.019 |
| Research & Professional+ | 2.26(0.92) | | 6.61, p = 0.000 |
| Related | 1.80(1.32 | | 2.35, p = 0.017 |

The open-ended responses echoed additional computational skills (e.g., computational methods for physics and chemistry applications) and other domain science topics (e.g., wind energy, power flows).

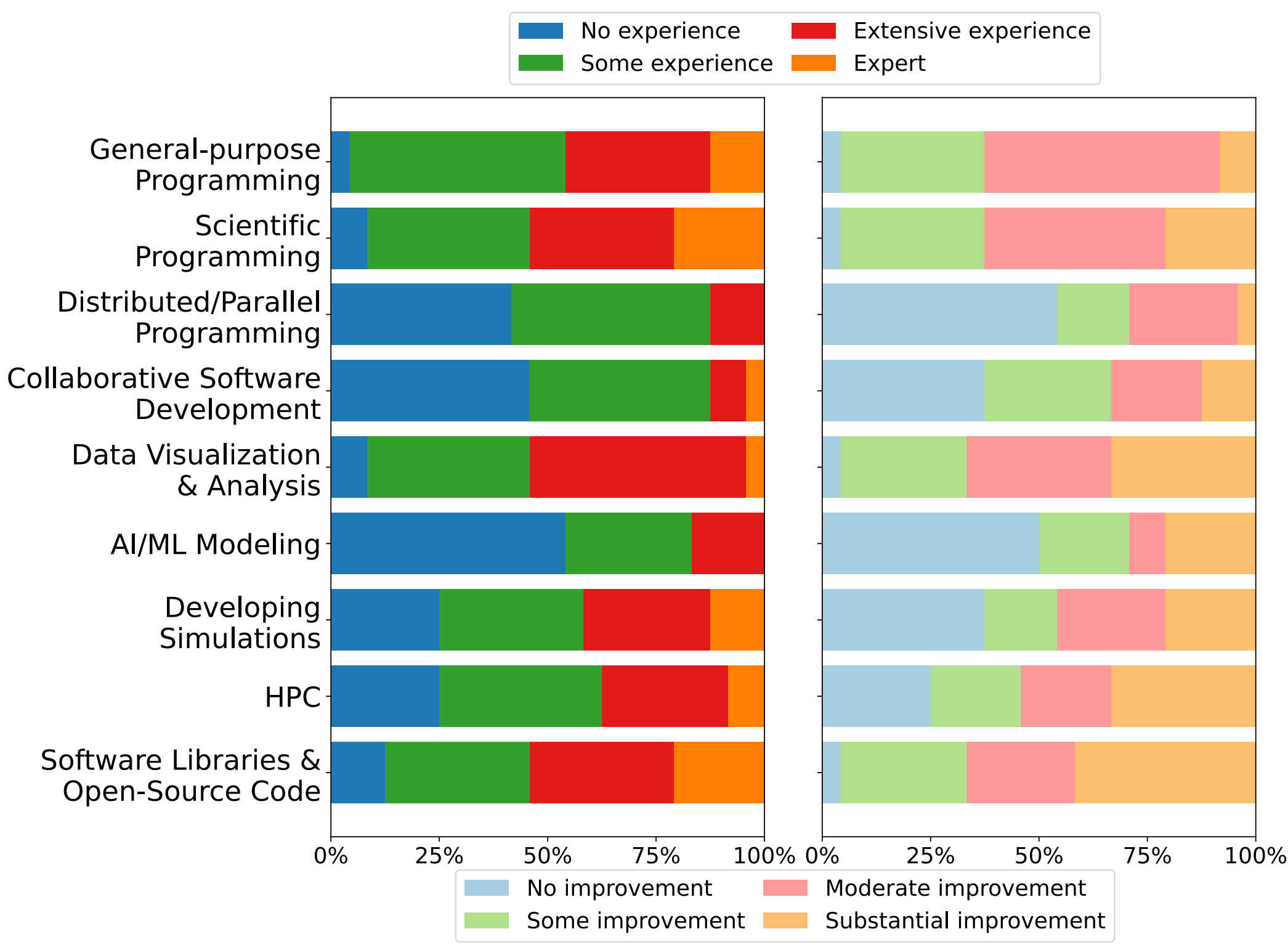

Figure 1. Participants' self-reported prior experience (top) and improvement of computational skills (bottom).

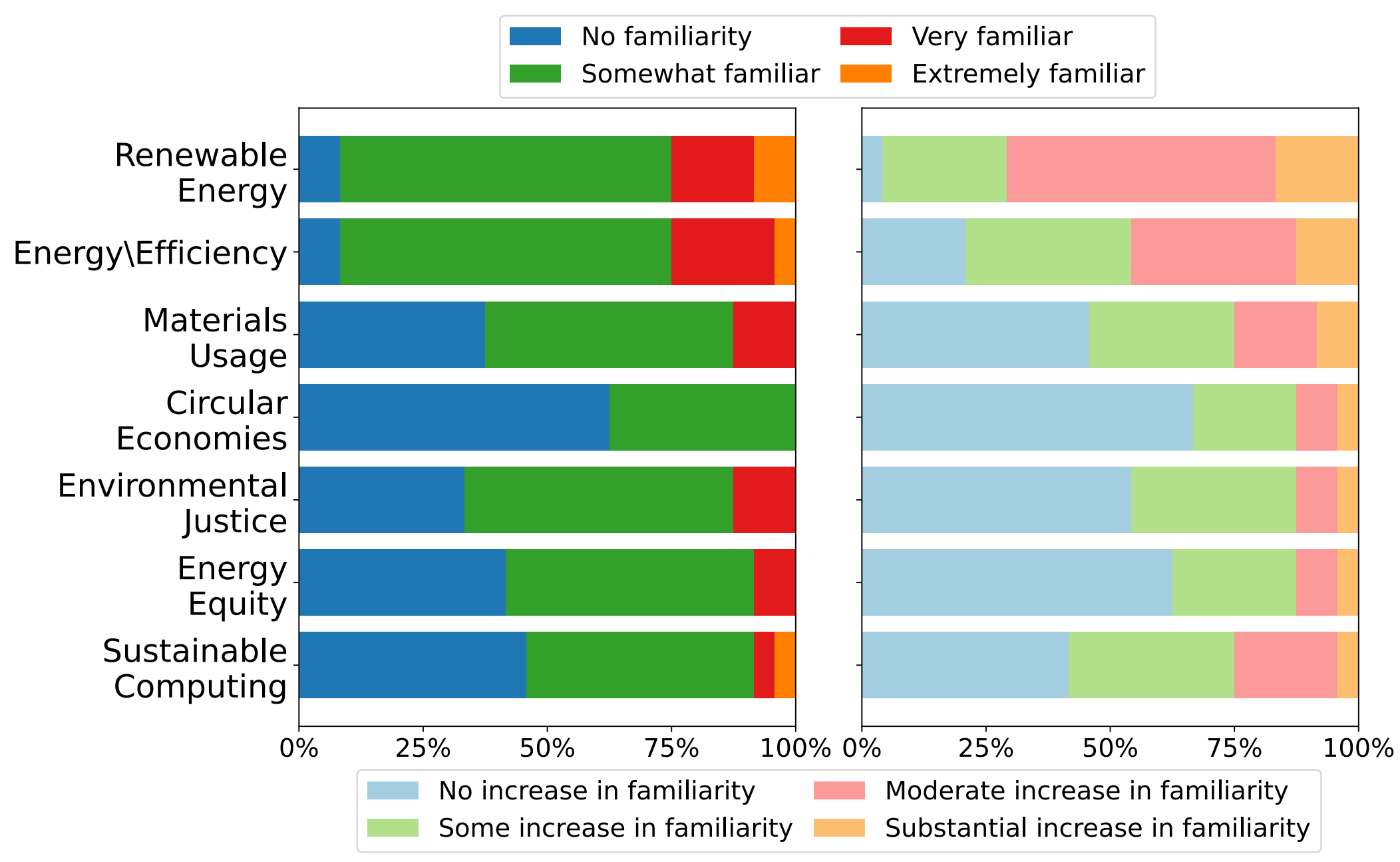

Fig. 2. Participants' self-reported prior familiarity (top) and improvement of familiarity (bottom) with sustainability and renewable energy topics.

*C. Research Skills*

On average, participants enter their internships with at least one prior research experience with respect to working on research projects, in research teams, and toward authoring publications. Additionally, participants report working on at least one research project ($t(23) = 2.94, p = 0.004$) or at least one research team during their internship ($t(23) = 2.41, p = 0.012$). The distribution of responses is provided in Figure 3.

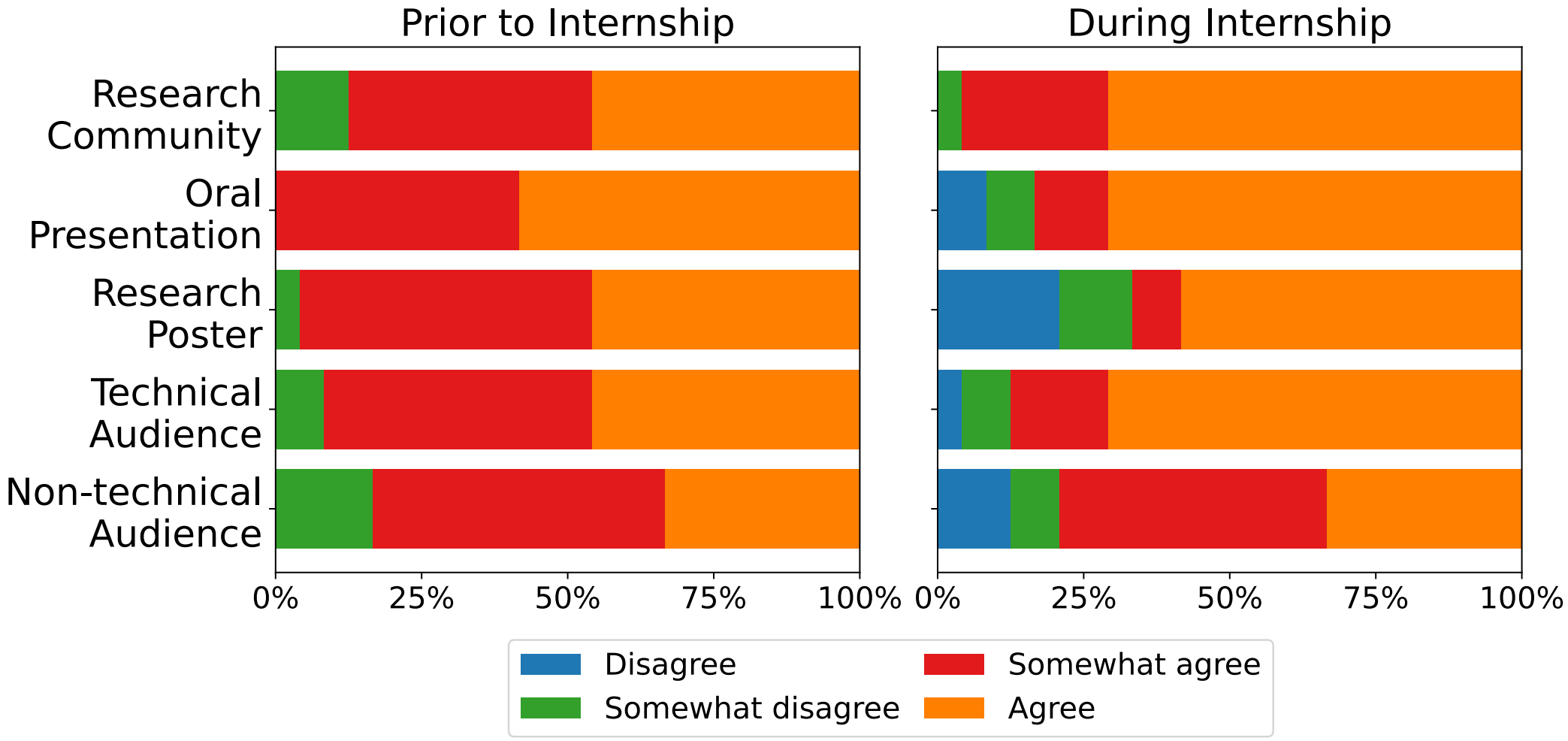

Fig 3. Participants' self-reported research experience before (left) and during (right) their internship.

The open-ended responses echoed similar experiences as our questions. For example, one participant noted "Remote collaboration" as a skill that improved, and another participant mentioned working in teams with varying levels of expertise as an important skill for their post-internship position.

*D. Professional Skills*

On average, participants enter their internships with the ability to communicate research across a variety of modes and are already part of research communities. Participants' agreement with the statement, "I'm part of a research community where I can get feedback on my work," is higher during their internship compared to before their internship ($t(23) = -1.88, p = .036$). While not statistically significant, we also note that more participants responded both "Agree" and "Disagree" with respect to these skills during their internship compared to before their internship. The distribution of responses is provided in Figure 4.

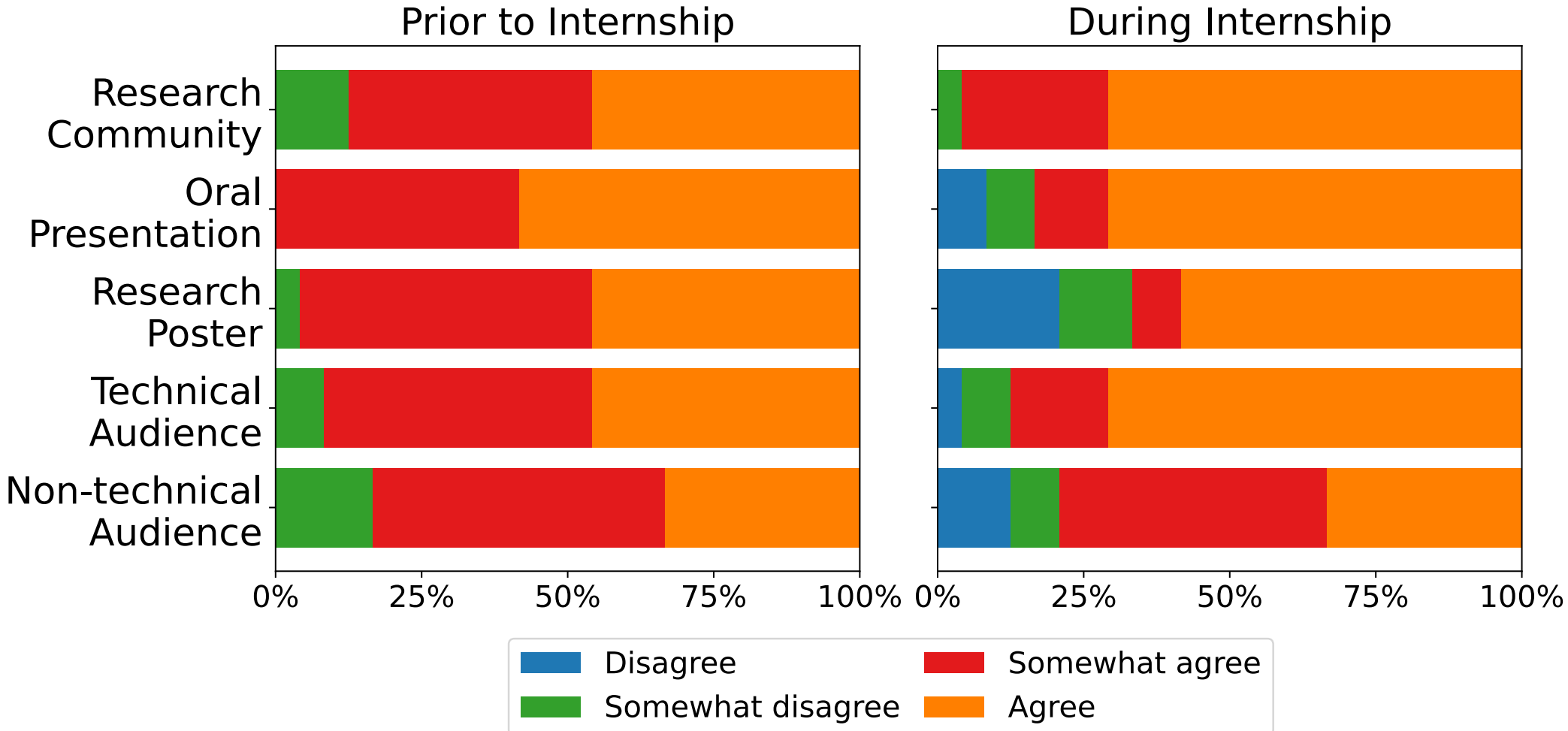

Fig. 4. Participants' self-reported agreement with being part of a research community and with ability to communicate ideas before (left) and during (right) their internship.

Five participants also mentioned networking or the networks they developed through their internships as professional skills in the open-ended responses.

On average, participants are interested across a variety of domains and sectors prior to their internships, except for nonprofits or other unspecified sectors. Additionally, interest across all domains increased after participants' internships, and no participants responded "Not interested" to any of the domains. Across sectors, participants were more interested in careers at NREL ($t(23) = -4.14, p = 0.000$) and other Department of Energy national laboratories ($t(23) = -2.60, p = 0.008$). The distribution of responses is provided in Figure 5 and Figure 6.

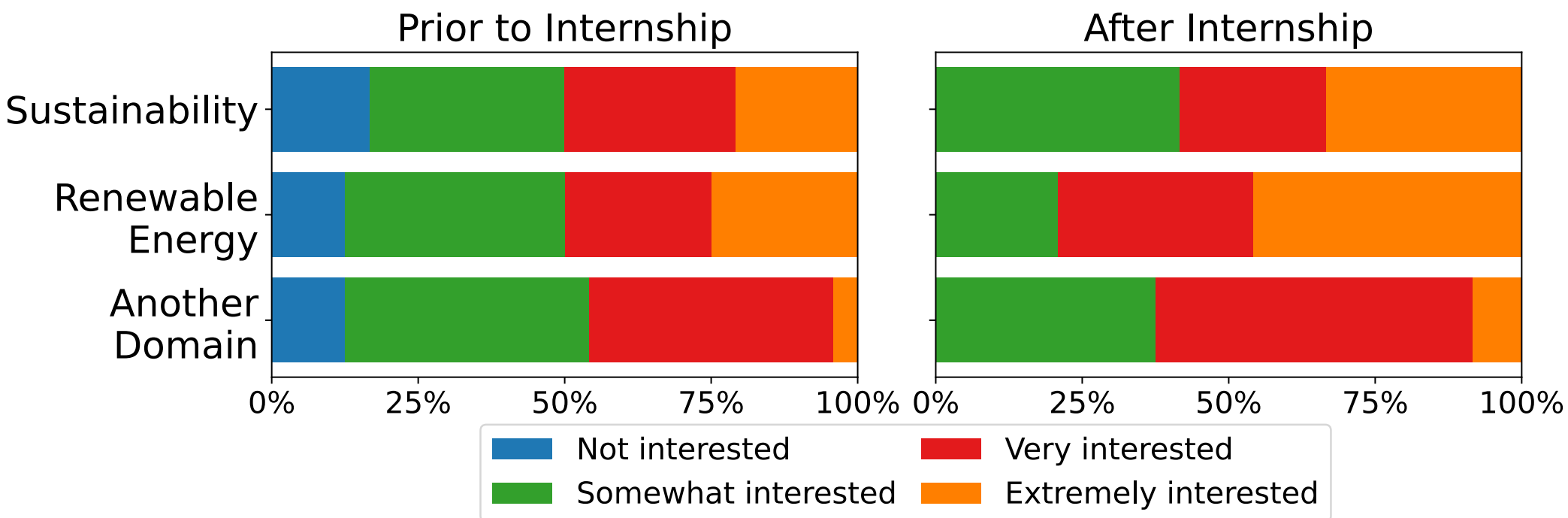

Fig. 5. Participants' self-reported interest in career domain before (left) and after (right) their internship.

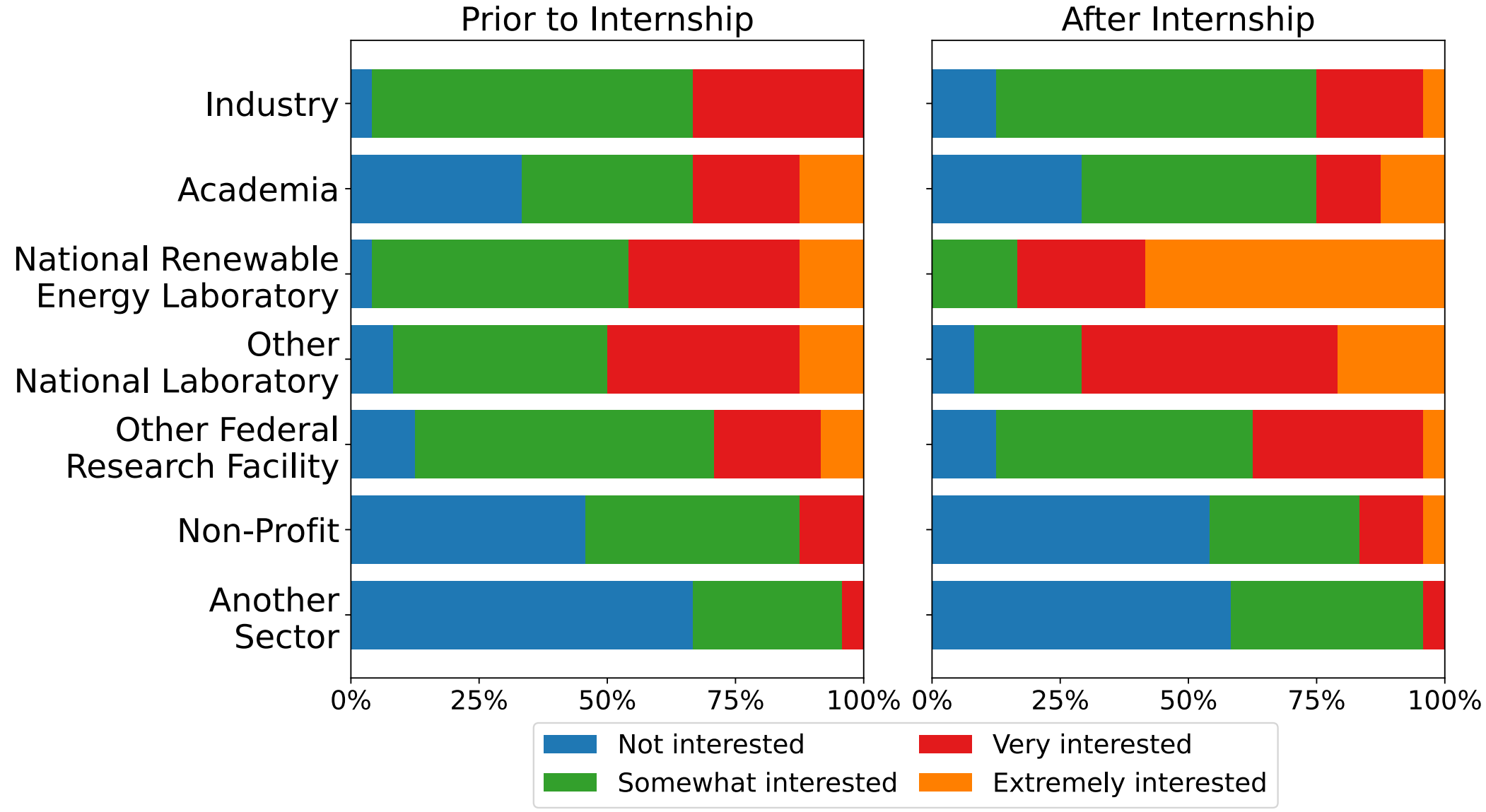

Fig. 6. Participants' self-reported interest in career sector before (left) and after (right) their internship.

## *E. Career Interests*

When asked about their current position at the time of completing the survey, 14 participants reported they were in the same or different degree program, and 10 participants reported they were looking for employment or currently employed. Of those in the same or different degree program, all participants reported enrollment in doctoral programs. Of those currently employed, participants reported working at DOE National Laboratories (including NREL), industry, academia, and government.

Out of the 15 participants who are not in the same degree program, only four participants reported they were looking for positions or currently in a field not related to NREL domain science topics. Additionally, participants on average reported skills improved during the internship were at least "Somewhat useful" in applying for positions.

Participants also reported skills improved during the internship were at least “Somewhat useful” in applying to their current position, and that they used these skills at least “Some of the time” in their current positions (see Table IV).

## V. DISCUSSION & CONCLUSION

Summarizing, federally funded graduate internships in CSE topics at NREL appear to have improved their computational skills the most, followed by increases in familiarity with domain science topics. The improvement in computational skills may speak to the many opportunities for HPC training offered by NREL. Additionally, interns were provided additional research experience and seemed more interested in career paths at national laboratories. After their internships, interns take the skills they developed with them back to their degree programs and into future positions. As a whole, we take these findings as indicators of the success of the internship programs at NREL.

At a larger level, our results echo other calls to encourage graduate student participation in internships to provide supplemental training and additional learning opportunities outside of formal degree programs [Murguía, Burton and Cao, 2022]. Our results emphasize the importance of federally-funded or other externally-funded programs in providing these internship opportunities that focus on educational objectives, rather than internships funded from research projects that may focus on hiring candidates whose skills best align with individual projects. This approach may be particularly valuable in CSE, where there is a large disparity in institutional resources and ability to offer cross-training between

To the best of our knowledge, this is the first study to evaluate graduate-level internship programs at the intersection of CSE and domain science. We encourage others to use our survey as part of evaluation efforts of other internship programs to see if our results generalize to other host institution and program contexts. We also suggest future work on the impacts of other models of using national laboratory capabilities to provide training in CSE and HPC, such as academic partnership programs.

## ACKNOWLEDGMENTS

This work was authored in part by the National Renewable Energy Laboratory (NREL), operated by Alliance for Sustainable Energy, LLC, for the U.S. Department of Energy (DOE) under Contract No. DE-AC36-08GO28308. This work was supported by the Laboratory Directed Research and Development (LDRD) Program at NREL. The views expressed in the article do not necessarily represent the views of the DOE or the U.S. Government. The U.S. Government retains and the publisher, by accepting the article for publication, acknowledges that the U.S. Government retains a nonexclusive, paid-up, irrevocable, worldwide license to publish or reproduce the published form of this work.

The authors would like to thank Addyson Sisemore, Jahi Simba, and Veronica Waller from NREL’s Human Resources team for their help in identifying potential participants based on internal records; Tony Magri from NREL’s IT Support team for his help in identifying interns who had worked NREL’s HPC resources. We would also like to thank Ken Oswald from the Oak Ridge Institute for Science and Education for providing the full list of NSF-MSGI recipients who completed their internship at NREL.